\documentclass[aps,prl,twocolumn,10pt,superscriptaddress]{revtex4-1}

\usepackage[latin9]{inputenc}
\usepackage{bm}
\usepackage{amssymb}
\usepackage{amsmath}
\usepackage{multirow}
\usepackage{graphicx}
\usepackage{breakurl}
\usepackage{float}
\usepackage{xcolor}
\usepackage{color}
\usepackage{hyperref}
\hypersetup{
    breaklinks=true,
	pdfnewwindow=true,      % links in new window
	colorlinks=true,       % false: boxed links; true: colored links
	linkcolor=blue,          % color of internal links (change box color with linkbordercolor)
	citecolor=blue,        % color of links to bibliography
	filecolor=blue,      % color of file links
	urlcolor=blue,          % color of external links
}

\newcommand {\B}[1]{\textcolor{blue}{#1}}

\begin{document}

\title{Bond ordering and phase transitions in Na$_{2}$IrO$_{3}$ under high pressure}

\author{Kaige Hu}
\affiliation{International Center for Quantum Materials, School of Physics, Peking
	University, Beijing 100871, China}
\affiliation{School of Physics and Optoelectronic Engineering, Guangdong University
	of Technology, Guangzhou 510006, China}
\author{Zhimou Zhou}
\affiliation{International Center for Quantum Materials, School of Physics, Peking
	University, Beijing 100871, China}
\author{Yi-Wen Wei}
\affiliation{International Center for Quantum Materials, School of Physics, Peking
	University, Beijing 100871, China}
\author{Chao-Kai Li}
\affiliation{International Center for Quantum Materials, School of Physics, Peking
	University, Beijing 100871, China}
\author{Ji Feng}
\email{jfeng11@pku.edu.cn}
\affiliation{International Center for Quantum Materials, School of Physics, Peking
	University, Beijing 100871, China}
\affiliation{Collaborative Innovation Center of Quantum Matter, Beijing 100871,
	China}
\affiliation{CAS Center for Excellence in Topological Quantum Computation, University of Chinese Academy of Sciences, Beijing 100190,
	China}

\date{\today}

\begin{abstract}
The Kitaev model of spin-1/2 on a honeycomb lattice supports degenerate topological ground states and may be useful in topological quantum computation. Na$_{2}$IrO$_{3}$ with honeycomb lattice of Ir ions have been extensively studied as candidates for the realization of the this model, due to the effective $J_{\text{eff}}=1/2$ low-energy excitations produced by spin-orbit and crystal-field effect. As the eventual realization of Kitaev model has remained evasive, it is highly desirable and challenging to tune the candidate materials toward such end. It is well known external pressure often leads to dramatic changes to the geometric and electronic structure of materials. In this work, the high pressure phase diagram of Na$_{2}$IrO$_{3}$ is examined by first-principles calculations. It is found that Na$_{2}$IrO$_{3}$ undergoes a sequence of structural and magnetic  phase transitions, from the magnetically ordered phase with space group $C2/m$ to two bond-ordered non-magnetic phases. The low-energy excitations in these high-pressure phases can be well described by the $J_{\text{eff}}=1/2$ states.
\end{abstract}

\maketitle
% insert suggested PACS numbers in braces on next line
% \pacs{75.10.Jm, 75.30.Et, 75.10.Kt, 75.25.-j}

% insert suggested keywords - APS authors don't need to do this
% \keywords{Na$_2$IrO$_3$, high pressure, phase transition}

%\maketitle must follow title, authors, abstract, \pacs, and \keywords

In recent years the Kitaev model~\cite{Kitaev2006ap}, an exactly solvable two-dimensional spin-$1/2$ model on a honeycomb lattice with distortional nearest-neighbor interactions, has attracted considerable attention.
The Kitaev ground state is a quantum spin liquid with possible non-abelian anyonic excitations, whose realization will be an important step towards topological quantum computation~\cite{Kitaev2006ap}.
To date,
5\emph{d} iridates $A_{2}$IrO$_{3}$(\emph{A}=Na, Li)~\cite{Modic2014ncomms,Takayama2015prl,Williams2016prb,Jackeli2009prl,Singh2010prb,Chaloupka2010prl,Liu2011prbr,
Lovesey2012jpcm,Choi2012prl,Ye2012prbr,Katukuri2014njp,Yamaji2014prl,Rau2014prl,Chun2015nphys,Hu2015prl,Hermann2018prbr,Hermann2017prb}
have been extensively studied for the realization of the Kitaev model.
However, the ground states of these materials are not the desired spin liquid, but all magnetically ordered~\cite{Singh2010prb,Liu2011prbr,Choi2012prl,Ye2012prbr,Lovesey2012jpcm,Chun2015nphys,Hu2015prl,Williams2016prb,Takayama2015prl,Modic2014ncomms}.
It turns out that the existence of Heisenberg interactions and also off-diagonal interactions plays an important role in determining the magnetic configuration of the ground state \cite{Yamaji2014prl,Rau2014prl,Hu2015prl}, and the magnetic configuration is sensitive to structure deviations \cite{Hu2015prl}.

Although the Kitaev model has not been realized in iridates, many studies indicate that the Kitaev terms are the dominant interactions and the systems are near the spin liquid region in the parameter phase diagrams~\cite{Hu2015prl,Takayama2015prl,Williams2016prb,Yamaji2014prl,Katukuri2014njp}.
Therefore, it may be possible to find new Kitaev materials whose parameters fortunately locate in the zone of the Kitaev spin liquid phase.
The recent studies on $\alpha$-RuCl$_{3}$, OsCl$_3$, Cu$_2$IrO$_3$ and H$_3$LiIr$_2$O$_6$ are progresses in this direction~\cite{Zhou2016prb,Sheng2017prbr,Abramchuk2017jacs,Kitagawa2018nature}.
In addition, the electronic states of these candidate materials can be changed dramatically by the application of external fields or pressure. Pressure can even introduce dramatic geometric changes to these materials. Therefore,  whether these candidate materials can be tuned to the Kitaev ground state will be an essential question.
To date, magnetic-field-induced quantum spin liquid phases are reported in $\alpha$-RuCl$_3$, however the experimental data cannot be reconciled with the behavior of the Kitaev spin liquid~\cite{Baek2017prl, Zheng2017prl}.
Pressure-induced melting of the magnetic order is observed in $\alpha$-RuCl$_3$ \cite{Wang2017prb,Cui2017prb}.
Structural and magnetic transitions under pressure are observed in iridate $\alpha$-Li$_{2}$IrO$_{3}$~\cite{Hermann2018prbr}.
As the first condensed matter candidate for the Kitaev model, however, Na$_{2}$IrO$_{3}$ shows no sign of structural phase transition in previous experiment under high pressure up to 24 GPa~\cite{Hermann2017prb}.

In this work, we study the phase transitions in Na$_{2}$IrO$_{3}$ under high pressure by first-principles calculations.
In Na$_{2}$IrO$_{3}$, each Ir$^{4+}$ ion is surrounded by an oxygen
octahedron and the crystal field splits $d$ orbitals into $e_g$ and $t_{2g}$ orbitals, and further the strong spin-orbit coupling (SOC) leads
to an effective pseudospin-$1/2$.
The interest in possible exotic quantum phase in Na$_{2}$IrO$_{3}$ is accompanied by many revisions of its structure, magnetic configuration, and microscopic model \cite{Singh2010prb,Chaloupka2010prl,Liu2011prbr,Lovesey2012jpcm,Choi2012prl,Ye2012prbr,Katukuri2014njp,Yamaji2014prl,
Rau2014prl,Chun2015nphys,Hu2015prl}.
It may reflect the complexity of 5\emph{d} transition metal oxides
due to the interplay of SOC, electron correlation, and crystal-field
splitting effects, all of which can be modified in non-trivial ways
by the application of external pressure.
Indeed, we find remarkable nonmagnetic (NM) ground states of Na$_{2}$IrO$_{3}$ under
high pressure, in which $J_{\text{eff}}=1/2$ states are still dominant
in the low-energy region. The magnetic phase transition is seen to
be induced by bond ordering, where local structure dimerization is
formed with long-range order, with a concomitant  electronic phase
transition from a Mott insulator to band insulators. The bond-ordered
non-magnetic phase may bear remarkable resemblance to the gapped A-phase
of the Kitaev ground state~\cite{Kitaev2006ap}.

We perform noncollinear relativistic density functional theory
calculations with full self consistent fields, as implemented in the
Vienna \emph{ab initio} simulation package (VASP)~\cite{Kresse1993prb,Kresse1996prb,Kresse1999prb}.
The projector-augmented wave potentials with a
plane-wave cutoff of 500 eV are employed.
We set $U=1.7$~eV \cite{Hu2015prl} and $J=0.6$ eV \cite{Marel1988prb}, corresponding
to a choice of effectively $U_{\text{eff}}=U-J=1.1$ eV \cite{Dudarev1998prb}.
The energy convergence criteria is $10^{-5}$ eV and the interatomic
force convergence for structure optimizations is $0.01$ eV/\AA.~Hydrostatic
pressures are adopted in our high-pressure study, i.e., pressures
are isotropic in all direction.
Phonon dispersions are calculated by the finite displacement method
\cite{Parlinski1997prl},
where a $2\times2\times2$ supercell is adopted.

\begin{figure}
	\centering
	\includegraphics[width=3.2in]{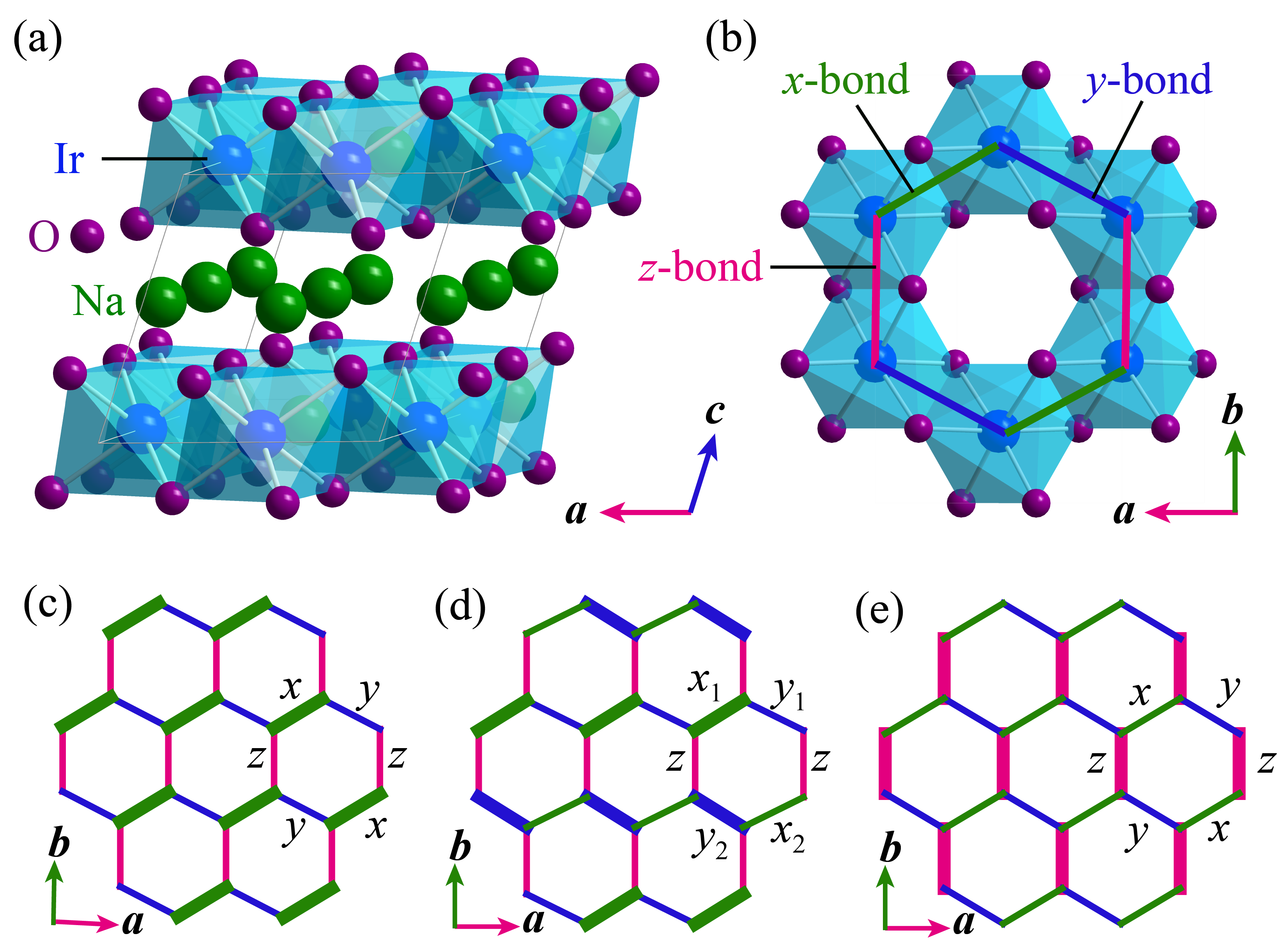}
	\caption{\label{lattice}(Color online) (a) Crystal structure of $C2/m$, viewed from slightly off the $\boldsymbol{b}$ direction.
		(b) Nearly ideal Ir honeycomb lattice of $C2/m$, viewed from the direction perpendicular to the \emph{ab} plane.
		Green, blue and magenta bonds are $x$-, $y$- and $z$-bonds, respectively. The bond length relation is $l_{x}=l_{y}\lesssim l_{z}$.
		The Ir honeycomb lattice of (c) $P\bar{1}$ with shorter \emph{x}-bonds: $l_{x}<l_{y}\approx l_{z}$,
		(d) $P2_{1}/m$ with shorter $x_{1}$- and $y_{2}$-bonds: $l_{x_{1}}=l_{y_{2}}<l_{x_{2}}=l_{y_{1}}\approx l_{z}$,
		and (e) $C2/m$-zz with shorter \emph{z}-bonds: $l_{z}<l_{x}=l_{y}$.
		Note that all short Ir-Ir bonds in (c-e) are highlighted by thicker lines.
}
\end{figure}

Under ambient pressure, Na$_{2}$IrO$_{3}$ is a layered compound
of space group $C2/m$ (No. 12) \cite{Choi2012prl,Ye2012prbr}, whose atom layers
are stacked repeating the sequence O-Ir$_{2/3}$Na$_{1/3}$-O-Na as
shown in Fig. \ref{lattice}(a). Each ion in the structure has six
oppositely charged ions as the first-neighbors forming octahedral
cages, akin to a distorted rock salt structure. Figure \ref{lattice}(a)
highlights the IrO$_{6}$ octahedrons particularly.
The atom layers are in the \emph{ab} plane. In each Ir$_{2/3}$Na$_{1/3}$
layer, Ir$^{4+}$ ions form an honeycomb lattice, with Na$^{+}$ ions
at the center of hexagons. There are three types of nearest-neighbor
Ir-Ir links named as $x$-, $y$-, and $z$-bonds (Fig. \ref{lattice}(b)),
respectively. This nomenclature is derived from the fact that those
bonds are perpendicular to the cubic $\mathbf{x}$, $\mathbf{y}$,
and $\mathbf{z}$ axes of the parent rock salt structure, respectively.
The Ir honeycomb lattice is nearly ideal, with \emph{z}-bonds slightly
longer than \emph{x}- and \emph{y}-bonds, i.e., $l_{x}=l_{y}\lesssim l_{z}$,
where $l_{x}$, $l_{y}$ and $l_{z}$ are bond lengths of \emph{x}-,
\emph{y}- and \emph{z}-bonds, respectively. The Ir honeycomb lattice
becomes zigzag antiferrmagnetic (AFM) below $T_{N}=15K$~\cite{Singh2010prb,Liu2011prbr,Choi2012prl,Ye2012prbr}
, and neighboring Ir layers are also
antiferromagnetically coupled \cite{Liu2011prbr}. We call this ground
structure $C2/m$-zigzag. Within the zigzag AFM phase, the direction
of magnetization is $\boldsymbol{g}\approx\boldsymbol{a}+\boldsymbol{c}$, located
in the cubic $xy$ plane of the IrO$_{6}$ octahedron and pointing
to the center of the O-O edge~\cite{Hu2015prl,Chun2015nphys}. Despite
the AFM stacking Ir honeycomb layers, only one Ir layer need to be
considered in the Kitaev-Heisenberg (KH) model since the interactions between Ir
honeycomb layers are negligible \cite{Hu2015prl}.

Under high pressure, two new crystal structures are found, whose space
groups are $P\bar{1}$ (No. 2) and $P2_{1}/m$ (No. 11), respectively.
Their structural stabilities are verified by phonon dispersions calculations (see the Supplemental Material~\cite{Hu2018sm}).
The key structural feature of these high-pressure phases is the emergence of bond ordering.
For $P\bar{1}$ shown in Fig.~\ref{lattice}(c), \emph{y}-bonds and
\emph{z}-bonds are elongated to the same extent whereas the $\emph{x}$-bonds
are shrunk distinctly, i.e., $l_{x}<l_{y}\approx l_{z}$. $P2_{1}/m$
shown in Fig.~\ref{lattice}(d) displays yet another bond ordering:
\emph{x}- and \emph{y}-bonds are further separated into two types\B{,}
i.e., $x_{1}$, $x_{2}$, and $y_{1}$, $y_{2}$, respectively, where
$z$-bonds, $x_{2}$-bonds, and $y_{1}$-bonds are elongated to some
extent while remaining almost equal to each other, but $x_{1}$-bonds
and $y_{2}$-bonds are shrunk, i.e., $l_{x_{1}}=l_{y_{2}}<l_{x_{2}}=l_{y_{1}}\approx l_{z}$.
It is worth mentioning that a third type of bond ordering shown in
Fig.~\ref{lattice}(e) are also possible as a metastable structure
at high pressures, where $z$-bonds are shrunk, i.e., $l_{z}<l_{x}=l_{y}$.
The space group remains $C2/m$, therefore we call this structure
$C2/m$-zz.
As shown in Figs. \ref{lattice}(c) and~\ref{lattice}(e), $C2/m$-zz  takes on
a similar bond ordering just as $P\bar{1}$.
We find $P\bar{1}$, $P2_{1}/m$ and $C2/m$-zz all  NM band insulators.

\begin{figure}
	%\centering
	\includegraphics[width=3.4in]{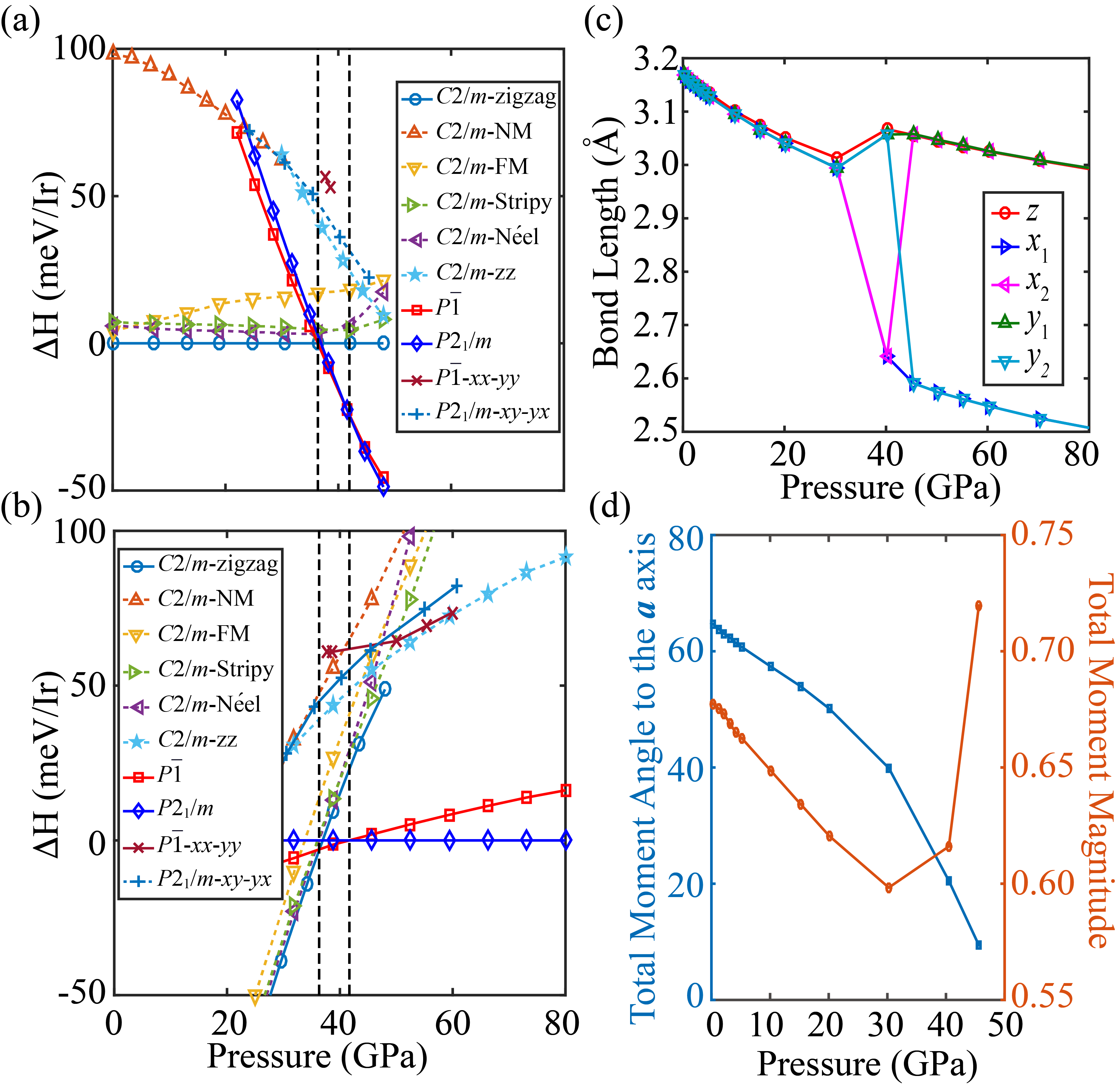}
	\caption{\label{bond_HP}(Color online) Calculated static lattice enthalpy of different structures relative to (a) the $C2/m$-zigzag structure and (b) the $P2_{1}/m$ structure, respectively.
		(c) Calculated bond lengths versus pressure, which reflects the changes of bond-ordering.
		(d) Calculated total moment magnitude and angle to the $\bm{a}$  axis of the $C2/m$-zigzag structure under various pressures.
	}
\end{figure}

According to our calculations, there is no structural phase transition below 24 GPa, which is consistent with previous experiment~\cite{Hermann2017prb}. However, at about 36 GPa, a structural phase transition is found computationally, and the new phase (space group $P\bar{1}$) is seen to be enthalpically favored up to 42 GPa. Since the $P\bar{1}$ structure is NM, there is simultaneously a magnetic phase transition at about 36 GPa.
At the same time, the system transforms from a Mott insulator to a band insulator.
At about 42 GPa, the computed enthalpies indicate the Na$_{2}$IrO$_{3}$ enters a new structural phase with space group $P2_{1}/m$, which remains NM.
Figures \ref{bond_HP}(a) and \ref{bond_HP}(b) show the enthalpy for $C2/m$, $P\bar{1}$ and $P2_{1}/m$ structures relative to the $C2/m$-zigzag structure and the $P2_{1}/m$ structure, respectively. A new energy reference is needed for higher pressures since the $C2/m$-zigzag structure can not exist under pressures higher than 48 GPa.
For completeness, bilayers cases of (1)
one layer with short $x$-bonds and the other layer with short $y$-bonds
and (2) one layer with short $x_{1}$- and $y_{2}$-bonds and the
other layer with short $x_{2}$- and $y_{1}$-bonds are also considered
and plotted, denoted as $xx$-$yy$ and $xy$-$yx$, respectively.
It turns out these bilayer structures are not energetically favored.

Figure \ref{bond_HP}(c) shows the changes of bond-ordering during phase
transitions by plotting bond lengths versus pressure. The bond lengths change dramatically between 35 GPa and 45 GPa, corresponding to two structural phase transtions discussed aboved. The bond-ordering
can be summarized as $l_{x}=l_{y}\lesssim l_{z}$ for $C2/m$-zigzag,
$l_{x}<l_{y}\approx l_{z}$ for $P\bar{1}$ and $l_{x_{1}}=l_{y_{2}}<l_{x_{2}}=l_{y_{1}}\approx l_{z}$
for $P2_{1}/m$, which is consistent with above discussions for Figs.
\ref{lattice}(c-e).

Previous finding suggests that the magnetic configuration is sensitive to structure deviations~\cite{Hu2015prl}. We then examine the magnetic structrure of the $C2/m$-zigzag structure during the structural phase transition under pressure.
Figure~\ref{bond_HP}(d) shows the total magnetic
moment magnitude and its angle to the $\bm{a}$ axis of the $C2/m$-zigzag structure under various pressures,
where the moment direction rotates in the \emph{ac} plane roughly
from $\boldsymbol{a}+\boldsymbol{c}$ to $\boldsymbol{a}$ as pressure increases.
The moment rotation is likely to be a consequence of the slight increase of the ratio $l_z/l_x$ as pressure increases, where the ratio trend is clearly shown in Fig.~\ref{bond_HP}(c).
As for the magnitude of magnetic moment, firstly it decreases gradually till about 30 GPa, and then goes up abruptly after 40 GPa, indicating a magnetic transition at around 30 $\sim$ 40 GPa, which is consistent with our previours discussions.

It should be remarked that $P\bar{1}$ and $P2_{1}/m$ do not appear
even metastable for $P<22$ GPa. $C2/m$-zigzag cannot
exist for pressures higher than about 48 GPa (Figs.~\ref{bond_HP}(a) and~\ref{bond_HP}(b)):
(1) for $48<P\leq56$ GPa, imaginary phonon frequency appears at $\Gamma$
(only $\Gamma$ is calculated for cost reasons), indicating the structure
unstability; (2) for $P>56$ GPa, $C2/m$-zigzag relaxes automatically
to $P\bar{1}$ or $P2_{1}/m$. Situations are similar for other $C2/m$
magnetic structures, i.e., they disappear above certain pressures
(Figs.~\ref{bond_HP}(a) and~\ref{bond_HP}(b)): (1) imaginary phonon frequency appears
at $\Gamma$ between 40 GPa and 50 GPa for FM, N\'eel  and NM, and between
50 GPa and 60 GPa for stripy; (2) they relaxes automatically to $P\bar{1}$
or $P2_{1}/m$ above a pressure between 60 GPa and 70 GPa for FM and
stripy, between 50 GPa and 60 GPa for N\'{e}el, and between 70 GPa and
80 GPa for NM.

\begin{figure}
	\includegraphics[width=74mm]{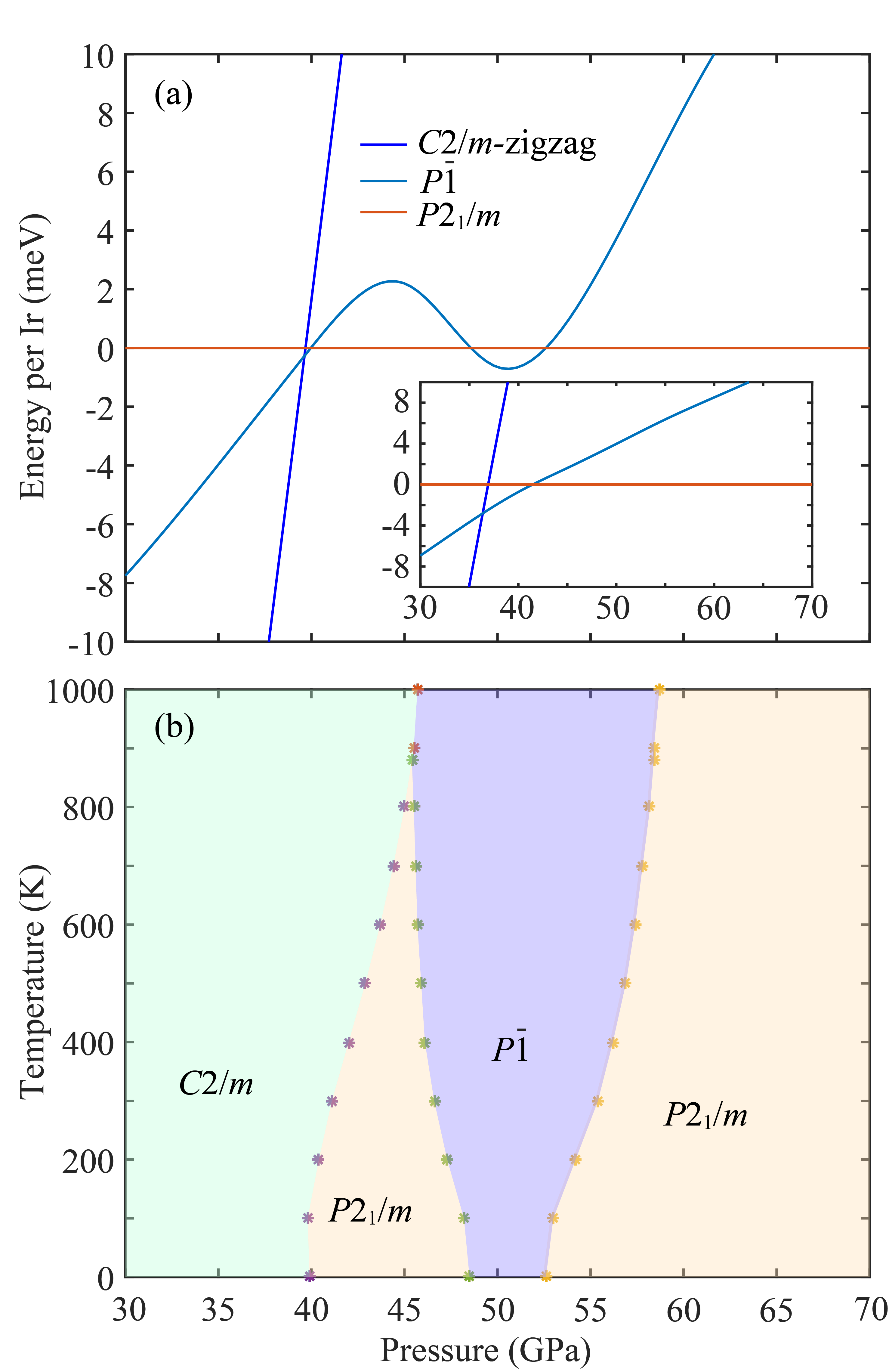}
	\centering
	\caption{\label{phase_diagram} (Color online) (a) The static lattice enthalpy of $C2/m$-zigzag, $P\bar{1}$ and $P2_{1}/m$ structures with zero point energy corrections. The inset in (a) shows the results without zero point energy corrections for comparision.
		(b) The pressure versus temperature phase diagram considering the phonon free energy.
	}
\end{figure}

The results presented so far, based on the static lattice approximation, already indicate the appearance of a rich phase diagram for Na$_{2}$IrO$_{3}$ under pressure.
We further take into account the effects of lattice vibrations by including the phonon free energy still within the Born-Oppenheimer approximation.
We calculate the phonon dispersions of $C2/m$-zigzag, $P\bar{1}$ and $P2_{1}/m$ under different pressures to obtain the zero point energy corrections (phonon free energy at zero temperature). The results are shown in Fig.~\ref{phase_diagram}(a). The inset in Fig.~\ref{phase_diagram}(a) shows the results without zero point energy corrections for comparision.
It can be seen that under pressures lower than 40 GPa, $C2/m$-zigzag is still the most stable structure.
For pressures higher than 53 GPa, $P2_{1}/m$ is the most stable. However, the stability range of pressure for $P\bar{1}$ shrinks to a point in contrast to $\sim$ 5 GPa without zero point energy corrections. In other words, with the increase of pressure, Na$_{2}$IrO$_{3}$ will first undergo a phase transition from $C2/m$-zigzag magnetic order to $P2_{1}/m$ NM order.
We also find that $P\bar{1}$ reenters as the most stable structure between 48 and 53 GPa. In Fig.~\ref{phase_diagram}(b), we plot the calculated phase diagram considering the phonon free energy. It is worth mentioning that in a relatively large temperature range, Na$_{2}$IrO$_{3}$ will undergo successive phase transitions, $C2/m \rightarrow P2_{1}/m \rightarrow P\bar{1} \rightarrow P2_{1}/m$, with the increase of pressure.

\begin{figure}
	\includegraphics[width=3.4in]{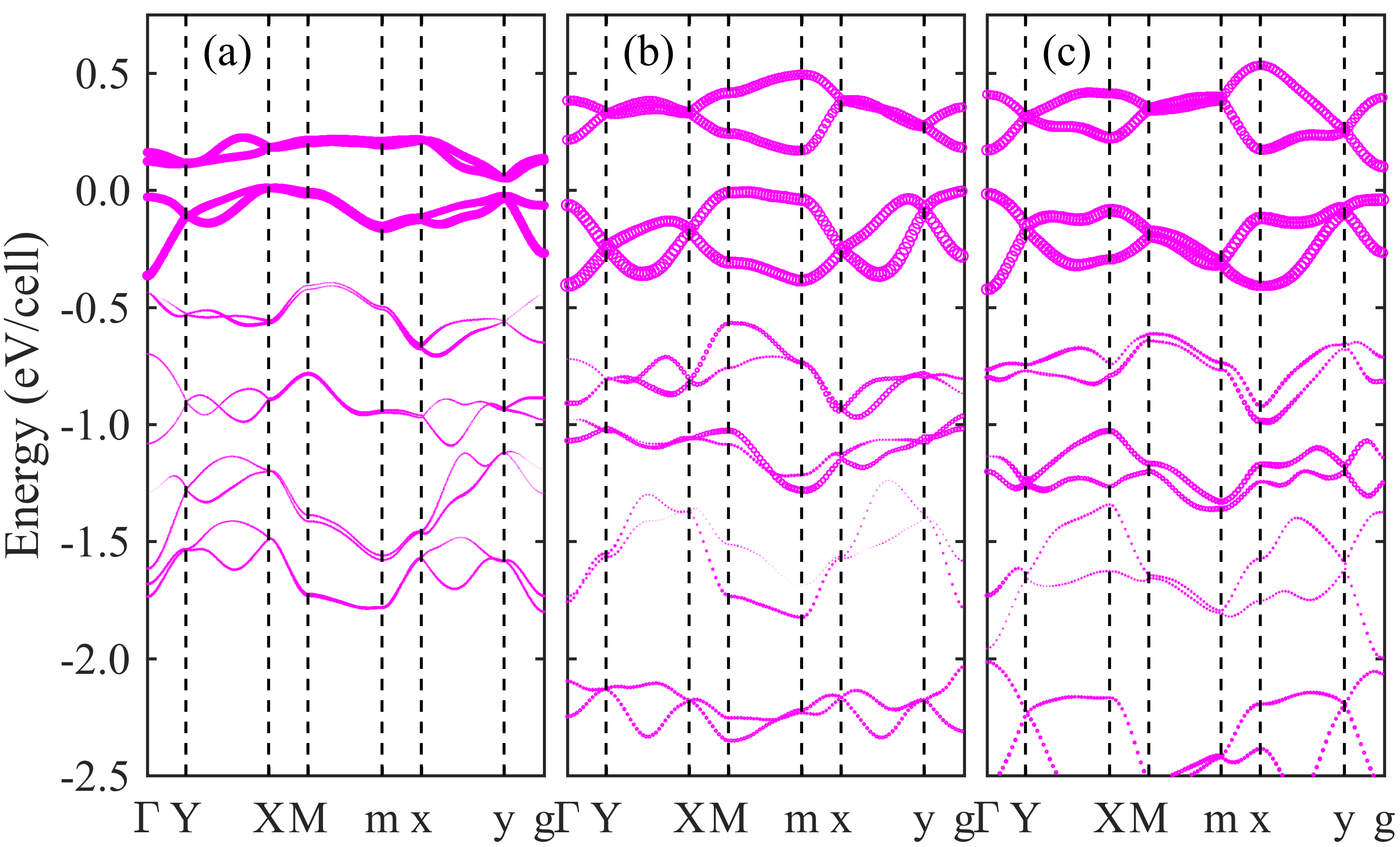} \caption{\label{TB_band}(Color online)
		Calculated tight-binding band structures (4 Ir per unit cell) with SOC of $C2/m$-zigzag (a), $P\bar{1}$ (b) and $P2_{1}/m$ (c) at 48 GPa, respectively.
		Fat-bands are plotted for band structures with SOC, where the line
		width represents the weight of the $J_{\text{eff}}=1/2$ states. The high symmetry $\mathbf{k}$-points are $\Gamma(0,0,0)$, Y$(0,1/2,0)$, X$(1/2,0,0)$, M$(1/2,1/2,0)$, m$(1/2,1/2,1/2)$, x$(1/2,0,1/2)$, y$(0,1/2,1/2)$, and g$(0,0,1/2)$.}
\end{figure}

In order to investigate the $J_{\text{eff}}=1/2$ feature of Na$_{2}$IrO$_{3}$
during phase transitions, we construct a first-principles based Wannier
tight-binding model~\cite{Mostofi2014cpc}. There are 4 Ir atoms in each unit cell. Each Ir$^{4+}$ ion has five 5$d$ electrons, occupying six $t_{2g}$ orbitals due to crystal field splitting, assuming the Ir-O octahedra remains regular during the crystal phase transitions. As a result of strong SOC, the six $t_{2g}$ orbitals are further separated into two manifolds with $J_{\text{eff}}=3/2$ and $J_{\text{eff}}=1/2$. We then decompose the band structures into $J_{\text{eff}}=3/2$ and $J_{\text{eff}}=1/2$ components. The results are plotted in fat-bands in Figs.~\ref{TB_band}
(a-c), where the line width represents the weight of the $J_{\text{eff}}=1/2$ states. It turns out that
$J_{\text{eff}}=1/2$ states are still the main components around the Fermi level. Also, the $J_{\text{eff}}=3/2$ states are fully filled and $J_{\text{eff}}=1/2$ states are half filled, which are consistent with our expectations.

In summary, we study the bond-ordering induced phase transitions in Na$_{2}$IrO$_{3}$ under high pressure by first-principles calculations.
We find that the Na$_{2}$IrO$_{3}$ crystal will undergo successive structural and magnetic phase transitions, $C2/m \rightarrow P2_{1}/m \rightarrow P\bar{1} \rightarrow P2_{1}/m$, with the increase of pressure,
where the $C2/m$ structure
holds a zigzag magnetic order, while $P2_{1}/m$  and $P\bar{1}$ are
all non-magnetic. $C2/m$ is classified to a Mott insulator, while $P2_{1}/m$  and $P\bar{1}$ are all band insulators. The low-energy excitations of these bond-ordered high pressure phases can be well described by the $J_{\text{eff}}=1/2$ states.
Considering that the $P\bar{1}$ phase possesses a $l_{x}<l_{y}\approx l_{z}$ bond-ordering as well as its $J_{\text{eff}}=1/2$ nature, we may expect that it bears remarkable resemblance to the gapped A-phase of the Kitaev ground state~\cite{Kitaev2006ap}.
The $P\bar{1}$ and $P2_{1}/m$ phases are also good instances of NM Kitaev-relevant phases.
Band structures and phonon dispersions are calculated for comparison to future experiments.
Together with previous high-pressure experiments on $\alpha$-Li$_{2}$IrO$_{3}$~\cite{Hermann2018prbr} and $\alpha$-RuCl$_3$~\cite{Wang2017prb,Cui2017prb}, we may infer that structural and magnetic transitions driven by external pressure are universal in these Kitaev candidate materials.
The bond-ordered phases found in our work may also exist in other candidates in certain range of pressure, which will enrich the phase diagrams of these materials.

\begin{acknowledgements}
We thank Fa Wang for helpful discussions.
Numerical simulations are performed on Tianhe-I Supercomputer System
in Tianjin. This work is supported by National Natural Science Foundation
of China (Grant Nos. 11725415 and 11647108), National Basic
Research Program of China (Grant Nos. 2016YFA0301004 and 2018YFA0305601), the Key Research Program of the Chinese Academy of Sciences (Grant No. XPDPB08-4),
and the 100 Talents Program for Young Scientists of Guangdong University
of Technology (Project No. 220413139).
\end{acknowledgements}

K.H. and Z.Z. contributed equally to this work.

\end{document}